\documentclass[twocolumn]{aastex62}

\newcommand{\eg}{e.g.,\ }

\newcommand{\kms}{km~s$^{-1}$}

\newcommand{\OI}{O~{\sc i}}

\newcommand{\CII}{C~{\sc ii}}

\newcommand{\MgII}{Mg~{\sc ii}}

\newcommand{\SiII}{Si~{\sc ii}}
\newcommand{\SiIII}{Si~{\sc iii}}

\newcommand{\SII}{S~{\sc ii}}
\newcommand{\CaII}{Ca~{\sc ii}}
\newcommand{\TiII}{Ti~{\sc ii}}

\newcommand{\FeII}{Fe~{\sc ii}}
\newcommand{\FeIII}{Fe~{\sc iii}}

\newcommand{\Nifs}{$^{56}$Ni}

\newcommand{\sBV}{s$_{BV}$}

\newcommand{\ab}{$\sim$}

\newcommand{\iBmax}{t$^{i-B}_{max}$}
\newcommand{\igmax}{t$^{i-g}_{max}$}
\newcommand{\sgr}{$s_{gr}$}

\usepackage{amsmath}
\graphicspath{{./}{figures/}}

\received{\today}
\revised{}
\accepted{}

\submitjournal{ApJL}

\shorttitle{Identifying extreme SNe Ia}
\shortauthors{Ashall et al.}

\begin{document}

\title{\textit{Carnegie Supernova Project-II: } 
 A new method to photometrically identify sub-types of extreme Type Ia Supernovae}

\correspondingauthor{Chris Ashall}
\email{Chris.Ashall24@gmail.com}

\author{C. Ashall}
\affil{Department of Physics, Florida State University, Tallahassee, FL 32306, USA}

\author{J. Lu}
\affil{Department of Physics, Florida State University, Tallahassee, FL 32306, USA}

\author{C. Burns}
\affiliation{Observatories of the Carnegie Institution for Science, 813 Santa Barbara St., Pasadena, CA 91101, USA}

\author{E. Y. Hsiao}
\affil{Department of Physics, Florida State University, Tallahassee, FL 32306, USA}

\author{M. Stritzinger}
\affil{Department of Physics and Astronomy, Aarhus University, 
Ny Munkegade 120, DK-8000 Aarhus C, Denmark}

\author{N. B. Suntzeff}
\affiliation{George P. and Cynthia Woods Mitchell Institute for Fundamental Physics \& Astronomy, Texas A\&M University, Department of Physics and Astronomy, 4242 TAMU, College Station, TX 77843}

\author{M. Phillips}
\affil{Carnegie Observatories, Las Campanas Observatory, 601 Casilla, La Serena, Chile}

\author{E. Baron}
\affiliation{Homer L. Dodge Department of Physics and Astronomy, University of Oklahoma, 440 W. Brooks, Rm 100, Norman, OK 73019-2061, USA}
\affiliation{Hamburger Sternwarte, Gojenbergsweg 112, D-21029 Hamburg, Germany}

\author{C. Contreras}
\affiliation{Carnegie Observatories, Las Campanas Observatory, 601 Casilla, La Serena, Chile}

\author{S. Davis}
\affiliation{Department of Physics, Florida State University, Tallahassee, FL 32306, USA}

\author{L. Galbany}
\affil{Departamento de F\'isica Te\'orica y del Cosmos, Universidad de Granada, E-18071 Granada, Spain.}

\author{P. Hoeflich}
\affil{Department of Physics, Florida State University, Tallahassee, FL 32306, USA}

\author{S. Holmbo}
\affil{Department of Physics and Astronomy, Aarhus University, Ny Munkegade 120, DK-8000 Aarhus C, Denmark}

\author{N. Morrell}
\affiliation{Carnegie Observatories, Las Campanas Observatory, 601 Casilla, La Serena, Chile}

\author{ E. Karamehmetoglu}
\affil{Department of Physics and Astronomy, Aarhus University, Ny Munkegade 120, DK-8000 Aarhus C, Denmark}

\author{K. Krisciunas}
\affiliation{George P. and Cynthia Woods Mitchell Institute for Fundamental Physics \& Astronomy, Texas A\&M University, Department of Physics and Astronomy, 4242 TAMU, College Station, TX 77843}

\author{S. Kumar}
\affiliation{Department of Physics, Florida State University, Tallahassee, FL 32306, USA}

\author{M. Shahbandeh}
\affiliation{Department of Physics, Florida State University, Tallahassee, FL 32306, USA}

\author{S. Uddin}
\affiliation{Observatories of the Carnegie Institution for Science, 813 Santa Barbara St., Pasadena, CA 91101, USA}

\begin{abstract}
We present a  new method to photometrically delineate between  various sub-types of type Ia supernovae (SNe Ia). 
Using the color-stretch parameters, \sBV\ or \sgr, and the time of $i$-band primary maximum relative to the $B$-band (\iBmax) or $g$-band (\igmax) maximum it is demonstrated that 2003fg-like,
1991bg-like, and 2002cx-like SNe~Ia can readily be identified. 
In the cases of these extreme SNe~Ia,  their primary $i$-band maximum occurs after the time of the $B$ or $g$ band maxima.
We suggest that the timing of the $i$-band maximum 
can reveal the physical state of the SN~Ia explosion as it traces: i) the speed of the recombination front of iron group elements in the ejecta, ii) the temperature evolution and rate of adiabatic cooling in the ejecta and,  iii) the presence of interaction with a stellar envelope.  This photometric sub-typing can be used in conjunction with other SNe analysis, such as the Branch diagram, to examine the physics and diversity of SNe Ia. The results here can also be used to screen out non-Ia SNe from cosmological samples that do not have complete spectroscopic typing. Finally, as future surveys like LSST create large databases of light curves of many objects this photometric identification can be used to readily identify and study the rates and bulk properties of peculiar SNe~Ia.

\end{abstract}


\section{Introduction} 
\label{sect:intro}
Type Ia supernovae (SNe~Ia) have revolutionized the study of an immense volume of the cosmos. To date they have been used
to map out the expansion rate of the Universe, providing the first observational evidence that it is accelerating \citep{Riess98,Perlmutter99}. 
 SNe~Ia also provide a  measure of the local Hubble constant \citep[\eg][]{Burns18}, constrain the dark energy equation-of-state parameter $w$ \citep[\eg][]{Scolnic18}, and  have the potential to map out the local distribution of dark matter through peculiar velocity studies \citep{Feindtet15}. Currently the use of SNe~Ia as standardizable candles is not limited by sample size, but a complex matrix of subtle systematic errors, where the lack of understanding of SNe~Ia physics and diversity contributes a major fraction of the total error budget \citep{Betoule14}.

We now have a myriad of varieties of SNe~Ia. 
Maximum light spectra have been used to classify normal SNe~Ia into four
sub-groups: core-normal, shallow silicon, cool, and broad line \citep{Branch06}. 
There are also many sub-types of SNe~Ia including;
i) the luminous 1991T-like SNe \citep{Filippenko92,Phillips92}, 
ii) the sub-luminous 1991bg-like SNe \citep{Filippenko91bg,Leibundgut93}, iii) the broad, but faint 2002cx-like SNe \citep{Li03,Foley13,Jha17}, iv) the over-luminous and possibly ``Super-Chandrasekhar" mass SNe~Ia,  2003fg-like SNe~Ia \citep{Howell06,Hicken07}, iv) the 2002ic-like SNe~Ia which show evidence of strong interaction with their circumstellar medium \citep{Hamuy03},  v) and the 2006bt-like SNe Ia, which have broad, slowly declining light curves but lack a prominent secondary maximum in the $i$~band \citep{Foley10}. 
Identifying the differences and diversity of SNe~Ia is key if we are to improve upon their effectiveness as a multi-purpose cosmological tool and understand their diversity and progenitor scenarios. 

Commonly, SNe~Ia are classified using maximum light spectra. However, with just one spectrum it is often not possible to firmly identify to which sub-type of SNe Ia
the object belongs.   Furthermore, spectra require significant exposure times and it is not possible to obtain spectra of all transients. 
Current and future surveys, such as ZTF and LSST,  scan the night sky with a daily cadence, and  obtain light curves of many more supernovae than can be followed spectroscopically with current infrastructure.

Previous work has used the SNe~Ia light curve fitter, SiFTO, in attempt to distinguish between SNe~Ia sub-classes \citep{Gonz14}. 
Here we use multi-band light curve observations from the \textit{Carnegie Supernova Project I \& II} (CSP I \& II) to present a new way to identify rare  SNe that likely have a thermonuclear origin. We use direct observational parameters and  concentrate on the $i$-band as it shows the largest diversity among SNe~Ia.


\section{Observational Sample}
\label{sect:sample}
The \textit{Carnegie Supernova Project}  I \& II obtained an unprecedented sample of over 300  SNe~Ia light curves on a stable and well-measured photometric system \citep{Contreras10,stritzinger11,Krisciunas17,Phillips19,Hsiao19}. 
Here, we consider a subset of these objects characterized by high-cadence $B$, $V$, $g$, and $i$-band light curve coverage. 
Photometric data from the CSP were obtained  with the Swope telescope at Las Campanas Observatory, and  reduced, calibrated and template subtracted following the procedures described in \citet{Krisciunas17} and \citet{Phillips19}.
Definitive light curve photometry from the CSP-I was published by \citet{Krisciunas17} and those from CSP-II will be published in the near future (Suntzeff et al., in prep).


The selection criteria for building the sub-sample examined in this work  include:  pre-maximum $BVi$-band photometry, a clearly discernible time of maximum light, and  photmetric coverage extending through 30~days past maximum. The normal SNe have an error on the time of maximum ($t_{\rm{max}}$[err]) less than 0.5\,days, other SNe have $t_{\rm{max}}$[err]$<$0.6\,days, except for the 03fg-like objects which have a $t_{\rm{max}}$[err]$<$2.5\,d. A larger error budget was set for 
the peculiar SNe to ensure the sample sizes were large enough.

Following these requirements when using the $B$-band we identify: 103 normal SNe~Ia, 4 1991T-like, 6 03fg-like,
11 1991bg-like, 5 2002cx-like, and 2 2006bt-like SNe. When using the $g$-band instead of the $B-$ band the sample size is smaller. There are 60 normal, 0 1991T-like, 10 1991bg-like, 3 2003fg-like, and 5 2002cx-like SNe. Additionally, in both samples  we have added 3 more  2003fg-like SNe~Ia from the literature, which are SNe~2006gz \citep{Hicken07}
, 2012dn \citep{Taubenberger19}, and ASASSN-15pz \citep{Chen19}.
All of the objects in the sample were spectroscopically classified to  determine their sub-type.

\begin{figure*}
\centering
 \includegraphics[width=1.0\textwidth]{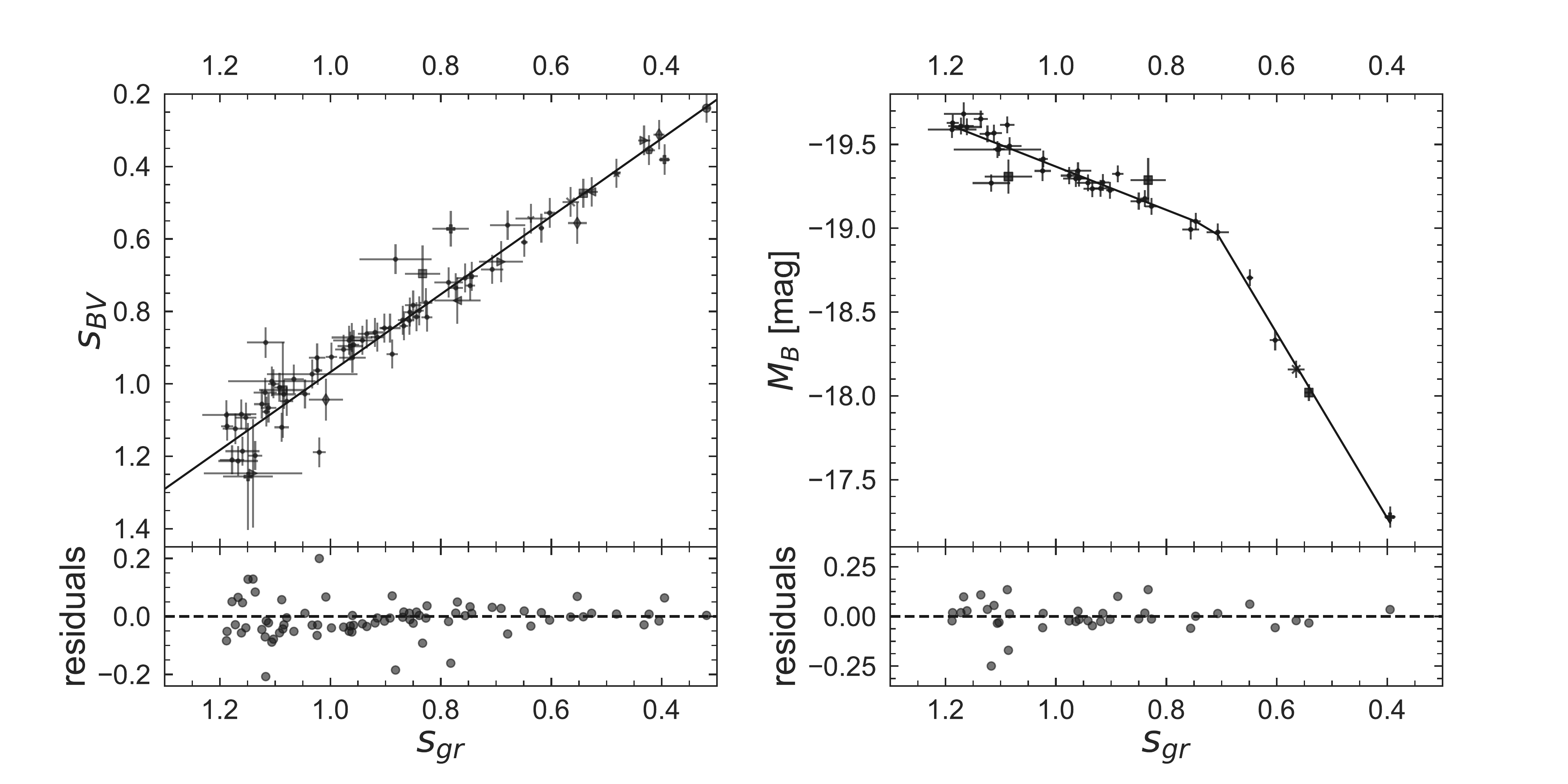}
\caption{\textit{Left:} \sBV\ vs \sgr\ produced using 79 SNe~Ia from the CSP I \& II. \textit{Right:} $M_{B}$ vs \sgr\ for a sample of 38 SNe~Ia which have $M_{B}$ values from \citet{Burns18}.  }
\label{fig:sbvsgr}
\end{figure*}


We have  corrected the SNe for Galactic extinction using values from \citet{Schlafly11}, but not for host-galaxy extinction. The average redshift of the SNe in the sample is 0.03$\pm$0.02. Therefore,  we have not  performed K-corrections because all of our objects are low redshift (less than 0.1), therefore the effect on the time of maximum will be minimal.  It is also non-trivial to compute K-corrections for peculiar SNe~Ia sub-types until more spectra are available.
 However, the  three non-CSP SNe were S-corrected to the natural CSP system, using the spectra of SN\,2009dc.  Finally, all of the light curves were converted to rest frame  by scaling the number of days since maximum by $\frac{1}{(1+z)}$.


The parameters used in this analysis are the time of  $i$-band maximum relative to $B$ or $g$ maximum (\iBmax, \igmax ) and the colour-stretch parameters \sBV\  and \sgr\ \citep{Burns14}.
The color-stretch parameter, \sBV, is  dimensionless and 
 defined as the time difference between $B$-band maximum and the reddest point in the post-maximum $B-V$  color curve divided by 30 days, where typical SNe Ia have \sBV $\approx$1.
The \sBV\ values of the SNe were obtained from Gaussian Processes interpolation to the light curves.  The Gaussian Processes interpolation accounts for uncertainties through its covariance function.
As recent surveys are often limited to  Sloan ($ugriz$) filters, and lack the more extensive information obtained with the traditional Johnson ($BV$) filters, we  use a  new parameter \sgr.  \sgr\ corresponds to  the time of ($g-r$) maximum relative to $g$-band maximum, normalized by 30~days. \sgr\ is fitted using the same approach as \sBV. 
Using 79 SNe Ia, we find  a precise linear correlation   between \sgr\ and \sBV. The correlation is  represented by:
\begin{equation}
s_{gr} = 1.03(01) + 0.93(02)*(s_{BV} - 1),
\end{equation}
with an   RMS = 0.05, see the left panel of Figure \ref{fig:sbvsgr}. \sgr\ contains  similar information as \sBV. This is demonstrated in the right hand panel of Figure \ref{fig:sbvsgr} where there is a strong correlation between $M_{B}$ and \sgr. This is similar to the trend seen between $M_{B}$ and \sBV\ in \citet{Burns18}. We find that a two-piece linear function fits the data best with a break at 0.71. The data were fit using the \textsc{python} package  \textsc{pwlf}, where a global optimization algorithm was used to determine the best break point location by solving least squares fits.  The best fit equations are given by 
\begin{equation}
M_B = \begin{cases} 
   -19.00 - 1.29\left(s_{gr} - 0.71\right)  \mbox{if } s_{gr} > 0.71 \\
   -19.00 - 5.51\left(s_{gr} - 0.71\right)  \mbox{if } s_{gr} < 0.71.
\end{cases}
\end{equation}

To determine \iBmax\ and \igmax, the  $B$, $g$, and $i$-band light curves were fit with Gaussian Processes, where the errors are once again accounted for in the covariance function. The fitting process was performed using the Gaussian Processes function in SNooPy \citep{Burns11}. The SNe data were directly fit.  No templates were used in the fitting procedure as they are not available for the extreme SNe~Ia, and using normal SNe Ia templates would bias the values obtained.

\section{Maximum light spectra}
\label{sect:spectra}
SNe~Ia are dominated by line opacity. Where the elements in the ejecta and their ionization states determines what is observed. Understanding which lines are present in each sub-type 
can provide clues about the diversity which is seen in the light curves. Hence, in this section we discuss the differences in the maximum light spectra for each sub-type.

The maximum light  spectra of all SNe~Ia are dominated by intermediate mass and iron group elements and  lack H and He features. All of the spectra prominently display  the iconic \SiII\ $\lambda$6355 feature. 
Therefore they all belong to the SNe~Ia category and originate from the thermonuclear explosion of a white dwarf.  
All the various sub-types follow those commonalities,  and each spectrum has its own peculiarities that set it apart from a CN (core-normal, \citealt{Branch06}) SN~Ia. Below we discuss some of the differences. 

At maximum light a normal SNe~Ia spectrum mainly displays doubly and singly ionized species. Specifically; \CaII\ $\lambda\lambda$3968, 3933, \SiII\ $\lambda$4130, \MgII\ $\lambda$4481, \SiIII\ $\lambda$4552, \FeII\ $\lambda$5196, \FeIII\  $\lambda$5156, \SII\ $\lambda$5453+$\lambda$5606, \SiII\ $\lambda \lambda$5978 6355, \OI\ $\lambda$7771 and  \CaII\ $\lambda\lambda$8498, 8542, 8662 \citep[\eg][]{Branch06,Ashall18}.

1991T-like SNe  are  brighter and hotter than normal SNe~Ia, with features associated with doubly ionized  \FeIII\ and \SiIII. Due to this higher temperature 1991T-like SNe have a weak \SiII\ $\lambda$6355 feature. This high ionization state has been suggested to be caused by a large amount of \Nifs\ in ejecta and  heating of the photosphere \citep{Phillips92,Filippenko92}.

SN~2003fg-like  SNe are events which are as bright, and usually brighter, than normal SNe Ia. However, the ionization state of these SNe  is much lower than that of a 1991T-like. They look remarkably like normal SNe~Ia, with the exception that they exhibit weak \CaII\ features \citep{Taubenberger19}. They also have two strong \CII\ features at $\lambda$6580 and $\lambda$7234,  which points to a large amount of unburnt material in the ejecta \citep{Howell06,Hicken07}. \citet{Hachinger12} demonstrated that the ionization state of the ejecta does not correspond to the luminosity of these events, and concluded that there must be some additional source of energy, for example interaction with H/He deficient material, such as a stellar envelope.

\begin{figure*}
\centering
\includegraphics[width=\textwidth]{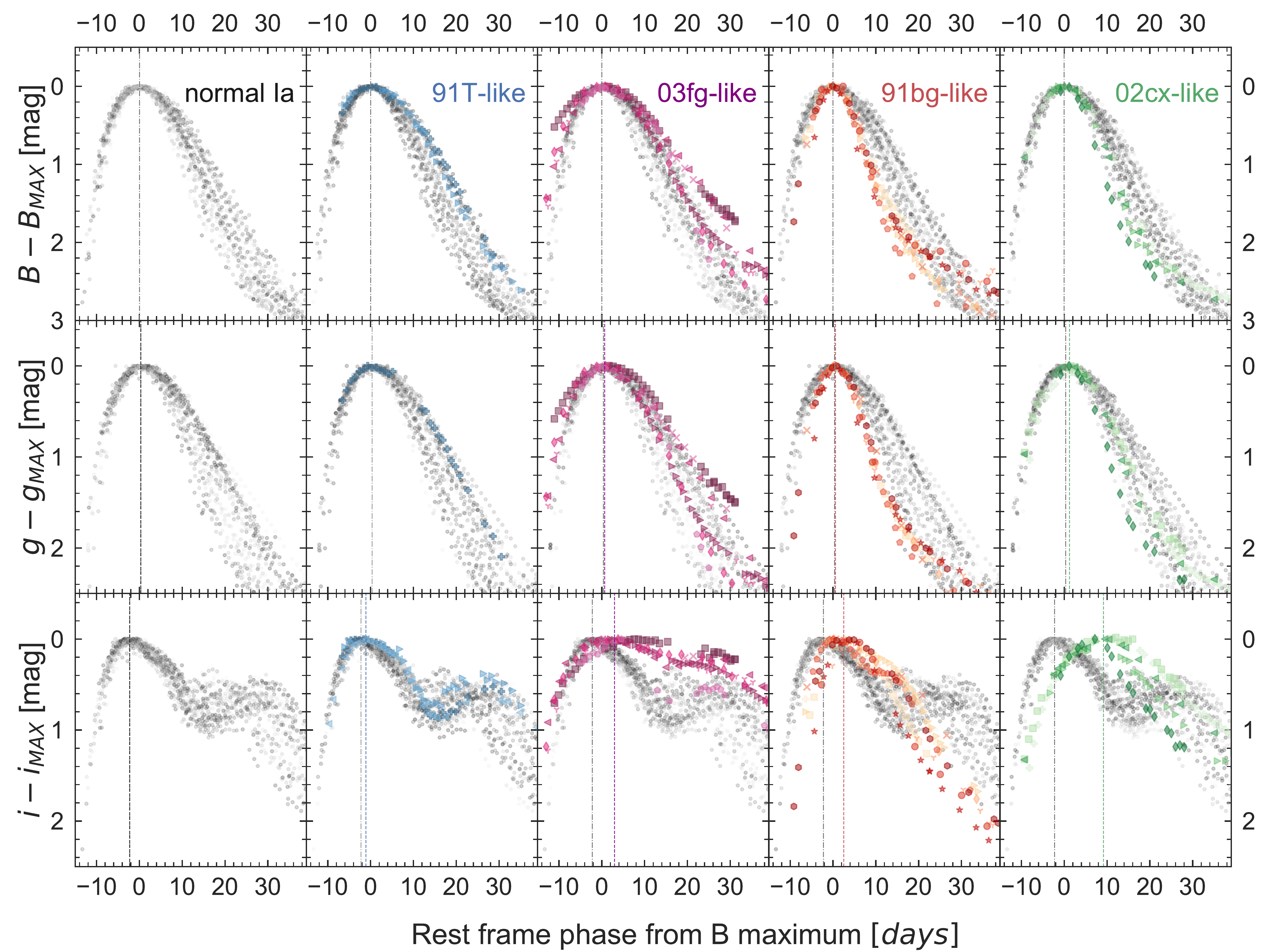}
\caption{ The $B$-band (top panels) $g$-band (middle panels) and  $i$-band (bottom panels) light curves for the SNe~Ia sub-types. For all of the panels the normal SNe~Ia are plotted in gray for comparison. For all panels the vertical gray dashed lines marks the mean time of maximum of the normal Ia population.
For the middle and lower panels the dashed colored lines mark the mean time of the  $g$-band (middle panels) or primary $i$-band (bottom panels) maximum for the corresponding sub-types.}
\label{fig:LCs}
\end{figure*}

2002cx-like are dominated by doubly ionized species, such as \FeIII, similar to that of 1991T-like events \citep{Li03,Phillips07,Foley13}. They are much less luminous and 
have low expansion velocities ranging from \ab2000--7000~\kms \citep{Jha17}. 

Finally, the spectra of 1991bg-like SNe are dominated by singly ionized species, a large \SiII\ line ratio, a strong \OI\ feature, and the appearance of a \TiII\ feature at 4400\AA\ \citep{Filippenko91bg,Leibundgut93}. It has been been suggested that this is due to  a temperature  evolution from normal to 1991bg-like SNe Ia, where the less luminous SNe have less \Nifs, less heating, and therefore a lower ionization state \citep{Nugent95}.


\section{Results}
\label{sect:LCs}
\subsection{Thermonuclear SNe}
In Fig.~\ref{fig:LCs} we present the $B$-, $g$-, and $i$-band light curves of the SNe in our sample separated by sub-type, normalized in both flux and the time of the $B$-band maximum. All of the normal SNe~Ia exhibit  a secondary $i$-band maximum  and the time of the primary $i$-band maximum (-2.25$\pm$0.91d) occurs prior to the epoch of $B$-band maximum.
In the $B$-band, 1991T-like SNe show a similar trend to the normal SNe Ia, but they tend to be broader than the normal SNe~Ia population. They also have a prominent  secondary $i$-band maximum, and a primary $i$-band maximum (-1.12$\pm$0.47d) that peaks before the $B$-band maximum. However, there is need of more 1991T-like high cadence light curves to determine the significant of this behavior. 

The 2003fg-like SNe~Ia may have longer rise times than normal SNe~Ia  and on average, they also have broader  $B$-band light curves. In the $i$~band they differ significantly from  normal objects. Their light curves are very broad and have a weak or no secondary maximum. Furthermore, the primary $i$-band maximum (2.98$\pm$2.10d) occurs after the time of $B$-band maximum.

The 1991bg-like SNe have $B$-band light curves that decline faster than normal SNe~Ia. In the $i$ band they also differ from  normal SNe~Ia  in that their primary maximum (2.51$\pm$1.61\,d) occurs after the time of $B$-band maximum. In the $B$-band, 2002cx-like SNe also decline rapidly, but less so than 1991bg-like SNe. In the $i$ band 2002cx-like SNe peak significantly later  (9.17$\pm$1.85\,d) compared to  all other sub-types. Both 1991bg-like and 2002cx-like SNe do not exhibit a prominent  $i$-band secondary maxima. 

 Similar trends are seen in the timing of $i$-band maximum relative to $g$-band maximum. For normal and 1991T-like SNe, the timing of the $i$-band primary maximum is before that of the $g$-band maximum, and for 1991bg-like, 2002cx-like, and 2003fg-like SNe Ia the $i$-band maximum is after that of the $g$-band maximum.

The $i$-band secondary maximum  is thought to be produced when the photospheric radius reaches its maximum; after which a recombination front of iron group elements recedes through the ejecta \citep{Hoeflich02,Kasen06,Jack15}. 
   For normal SNe the opacity drop starts at about +20 days, therefore they have a clear secondary $i$-band maximum. 
  This timing of the secondary $i$-band maximum  is a function of luminosity, brighter SNe~Ia have a later secondary $i$-band maximum and later recombination of iron group elements due to increased opacity from higher temperatures and larger \Nifs\ masses synthesized in the explosion. For 1991bg-like SNe, less luminosity implies less heating and an earlier drop in the opacities. The opacity drop starts  at  maximum  light and leads to the merging of the first and secondary $i$-band  maxima.  Hence 1991bg-like SNe  all have \iBmax$>0$.



\begin{figure*}
\centering
\includegraphics[width=\textwidth]{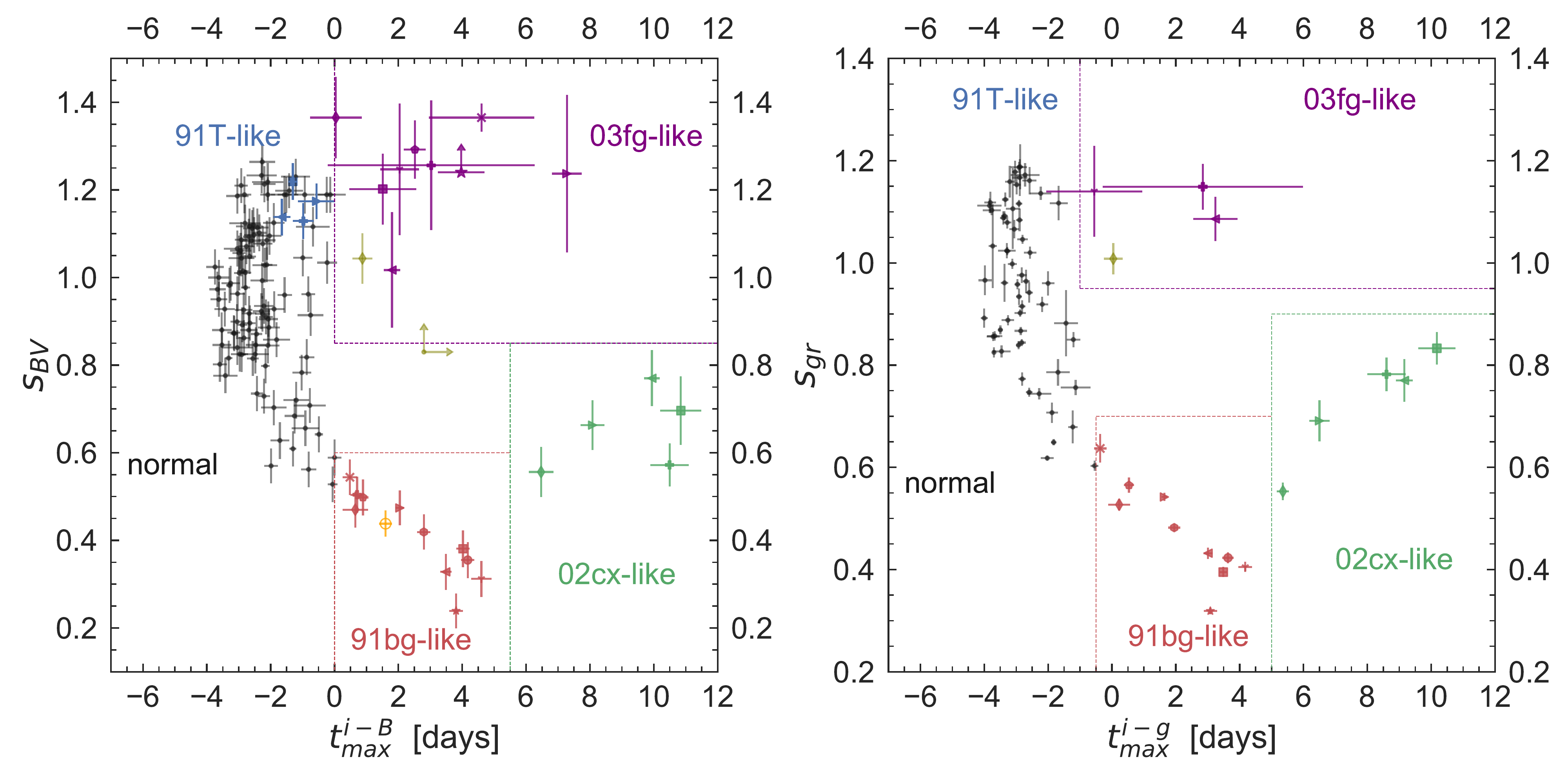}
\caption{ The color stretch parameter as a function of $i$-band maximum. \textit{Left:} Using \sBV\ and \iBmax\ there is a clear separation between sub-types. \textit{Right:} Using \sgr\ and \igmax\ the separation is also seen but the sample size is smaller. The mean values of the sample and parameters for the grouping can be found in Table \ref{table:Timax}. In the left panel SN~2006bt (yellow diamond) and upper limits for SN~2006ot (yellow circle) are also plotted.  SN~2006ot is a 2006bt-like SNe. For comparison we have also added SN~2016hnk (open orange circle) from \citet{Galbany19} in the left panel.   }
\label{fig:classplot}
\end{figure*}

The 2003fg-like SNe  have \iBmax \,$>0$, and ionization states similar or lower than normal SNe~Ia. Generally, they also show expansion velocities which are low at maximum light. For example the velocity of the \SiII\ $\lambda$6355 feature was \ab7500~\kms\ at maximum light for SN~2009dc \citep{Taubenberger11}.  These objects do not show the $H$-band break in the NIR spectra at +10\,d, which occurs when the photosphere reaches the \Nifs\ rich region \citep{hk96,Wheeler98,Hsiao15,Ashall19a,Ashall19b}. The variety of $H$-band break for different sub-types of SNe Ia can be seem in Fig. 6 from \citet{Hsiao19}. All of this may be an indication that these objects have a large mass of carbon, oxygen, or intermediate mass elements above the \Nifs\ layers.
This envelope  can effectively trap the gamma-rays  produced from the \Nifs,  leading to longer diffusion time scales and the recombination of iron group elements in the ejecta occurring over longer time scales.

The 2002cx-like SNe  have high ionization states but  $i$-band light curves that peak significantly later than normal SNe~Ia. 
2002cx-like SNe have the lowest expansion velocities of all the SNe~Ia, in the range of  2000-7000~\kms, which may cause them to stay hot and highly ionized for a long time. 
It has been  suggested that 2002cx-like SNe Ia  
come from a Chandrasekhar mass white dwarf progenitor that experiences a pulsational delayed detonation \citep{Stritzinger15}. In this scenario
the photosphere recedes through the ejecta quickly and only keeps the material just above the photosphere highly ionized. This produces the narrow lines, high ionization and lack of recombination of iron group elements leading to one $i$-band maximum. 
Alternatively, it has  been suggested that 2002cx-like SNe come from the partial deflagration of a white dwarf, where the ejecta is fully mixed. In this scenario the \Nifs\ in the outer layers produces the high ionization and a lack of recombination of iron group elements  \citep{Jha17}. 
Generally, the low expansion velocities, high ionization state, and broad light curves of these objects suggest that they have the longest diffusion time scales, which produces a slow temperature evolution and adiabatic cooling and is the cause of the large \iBmax.

The color stretch parameter, \sBV,  has successfully been used in conjunction with other observables to distinguish SNe~Ia properties \citep{Ashall18,Burns18}. 
Fig.~\ref{fig:classplot} shows  \sBV\ and \sgr\ as a function of \iBmax\ or \igmax. All of the extreme SNe, except 1991T-like, are classified by having \iBmax$>0$\,d. They are also all located in distinct areas of this diagram \footnote{It should be noted that this separation does not exist if using \sgr\ and the time of $g$-band maximum relative to $r$-band maximum.}. Table~\ref{table:Timax} contains the parameters that can be used to identify these groups. 


The 1991bg-like, 2003fg-like and 2002cx-like SNe are  separated from the normal population, which points to different origins of these subclasses. It is worth addressing if there is a continuous distribution between most of the SNe~Ia sub-types. For example, previous work has shown 1991bg-like SNe and normal SNe~Ia come from a continuous distribution \citep{Burns18,Ashall18}. This is also seen in Fig. \ref{fig:classplot} where 1991bg-like SNe are located at the end of the distribution of normal SNe~Ia. In fact, it could be argued that the division between 1991bg-like and normal SNe~Ia is slightly arbitrary and purely determined by the presence of \TiII\ and the ionization state seen in maximum light spectra.
SN 2003fg-like objects are often thought to be their own distinct sub-type, and from Fig. \ref{fig:classplot} this appears to be true. The peculiar SN~2006bt has been suggested to be a connection between 2003fg-like SNe and normal SNe Ia \citep{Foley10}, and its location in Fig. \ref{fig:classplot} fits with this conclusion. Finally, it seems that the 
2002cx-like SNe are a distinct group, and possibly come from a totally different origin.

\subsection{Core collapse SNe}
The aim of this work is to produce a photometric sub-typing of SNe~Ia, similar to that found with maximum light spectra in \citet{Branch06}. However, it is also interesting to  understand where core-collapse SNe are located in this parameter space, and to see if their location offers insights on the physics of the Ia subgroups. SNe~II  evolve over longer time scales than SNe~Ia. For example SN~2012aw, a type II-SN,  did not peak in $B-V$ until 90\,days past $B$-band maximum, which would give it  an \sBV\ of 3. 
The type II-L SN\,2013ej peaked in the $i$~band at least 15\,days after $B$-band maximum.  This places  SNe~II outside of both axes  of Fig.~\ref{fig:classplot}. We also examined a type IIb (SN~1993J, \citealt{Schmidt93}) which has \sBV\ of 0.9 and \iBmax$\approx$3, a Type Ib (SN~2009jf \citealt{Valenti11}) which has \sBV$\approx$0.91 and \iBmax$\approx$8, and a type Ic (SN~2016coi \citealt{Prentice18}), that has \sBV$\approx$0.59 and \iBmax$\approx$7. This places the Type IIb and Ib in the 2003fg-like area of Figure \ref{fig:classplot}, and the Type Ic supernova in the 2002cx-like area of the Figure \ref{fig:classplot}.

\begin{figure}
\centering
 \includegraphics[width=0.48\textwidth]{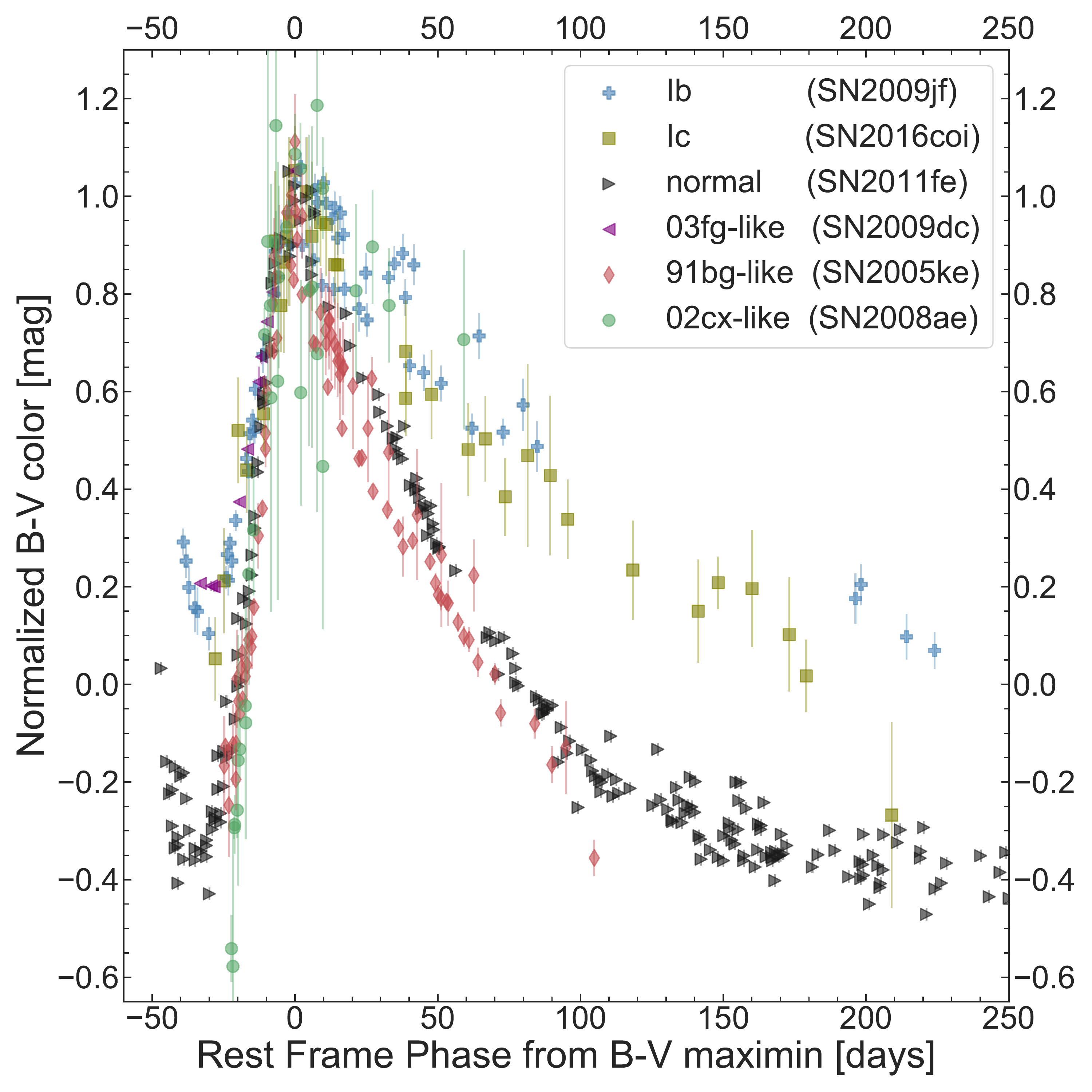}
\caption{
The $B-V$ colour curves of thermonuclear and core-collapse SNe as a function of time from $B-V$ maximum, normalised in both time and magnitude. The normal Ia, 1991bg-like SNe and 2002cx-like shown here all have a faster evolution after the time of $B-V$ maximum, where as the core-collapse SNe and 2003fg-like SNe have a slower evolution.}
\label{fig:color}
\end{figure}

Some additional information which could lead to clues about the SNe~Ia sub-types is the gradient of the $B-V$ or $g-r$ color curve after the color turnover. For normal, 1991bg-like, and some 2002cx-like SNe (e.g. SN~2008ae)  the slope  after the $B-V$ maximum is much steeper than for core collapse SNe. For example it takes SN\,2011fe 80\,days to change by  one magnitude, whereas it takes 150~days for the type Ic SN~2016coi to change by the same amount, see Fig. \ref{fig:color}.    However, the gradients of the color curves of core collapse supernovae and 2003fg-like SNe are similar\footnote{It should be noted that the gradient of the color curve is also sensitive to the  total-to-selective extinction ($R_{V}$).}.  This may indicate that the photosphere is moving through ejecta with similar composition, probably a large carbon/oxygen rich layer.  For the 2003fg-like SNe the interaction with this large H/He deficient envelope could also be the cause of the  excess luminosity compared to normal SNe~Ia \citep{Hachinger12}. 

Although  not the aim of this work, the results here may also be used alongside absolute magnitude to photometrically classify transient sources. For example, most stripped-envelope supernovae reach peak $B$-band absolute magnitudes of $M_B \approx -17$ to $-18$  \citep[see][their Fig.~8]{Taddia18}. However, even the faintest 2003fg-like SNe from the CSP sample and  the literature peak at --19.3\,mag \citep{Taubenberger19}, making 2003fg-like events consistently at least one magnitude more luminous. Therefore, a cutoff of $M_B = -18.5$\,mag could be used to separate core-collapse and 2003fg-like events. It should be noted that a larger sample of core collapse SNe needs to be examined to test the validity of these results. Also, further data of extreme SNe~Ia will enable the robustness of this method to be determined.

\begin{deluxetable*}{ccccccc}
\tablewidth{\textwidth}
\tablecaption{The average values of \sBV, \sgr, \iBmax, \igmax\ and the parameters used to classify the objects in Figure \ref{fig:classplot}.   \label{table:Timax}}
\tablehead{
\colhead{Sub-type}&
\colhead{mean(\sBV)}&
\colhead{mean(\sgr)}&
\colhead{mean(\iBmax)}&
\colhead{mean(\igmax})&
\colhead{Classification ($B$)}&
\colhead{Classification ($g$)}}
\startdata
 \hline
    normal	&	0.95	$\pm$	0.17	&	0.96	$\pm$	0.16	&	$-$2.25	$\pm$	0.91	&	$-$2.81	$\pm$	0.76	&	\iBmax$<0$ and \sBV$>$0.5	&	\igmax$<-0.5$ and \sgr$>$0.6	\\
    91T-like	&	1.17	$\pm$	0.04	&	$\cdots$	&	$-$1.12	$\pm$	0.47	&	$\cdots$		&	\iBmax$<0$ and \sBV$>$1.1	&	$\cdots$	\\
    03fg-like	&	1.25	$\pm$	0.11	&	1.13	$\pm$	0.03	&	2.98	$\pm$	2.10	&	1.84	$\pm$	2.09	&	\iBmax$>0$ and \sBV$>$0.85	&	\igmax$>-1$ and \sgr$>$0.95	\\
    91bg-like	&	0.41	$\pm$	0.10	&	0.47	$\pm$	0.10	&	2.51	$\pm$	1.61	&	2.14	$\pm$	1.59	&	$0<$\iBmax$<5.5$ and \sBV$<$0.6 	&	$-0.5<$\igmax$<5.0$ and \sgr$<$0.7	\\
    02cx-like	&	0.65	$\pm$	0.09	&	0.73	$\pm$	0.11	&	9.17	$\pm$	1.85	&	7.96	$\pm$	1.98	&	\iBmax$>5.5$ and \sBV$<$0.85	&	\igmax$>5.0$ and \sgr$<$0.9	\\
\enddata
\end{deluxetable*}

\section{Conclusions}
\label{sect:conclusion}
We present a new method to photometrically distinguish between sub-types of thermonuclear SNe.   A set of  $B$-, $g$-, and $i$-band light curves of 103 normal SNe~Ia, 4 1991T-like, 9 2003fg-like, 11 1991bg-like,  5 2002cx-like SNe, and 2 2006bt-like SNe have been  used to demonstrate how the timing of the primary $i$-band maximum provides a clear means to discriminate between SN sub-types.   


The peculiar 2003fg-like, 1991bg-like,  and 2002cx-like SNe have their primary peak in the $i$-band later than that of the $B$-band and $g$-band. These sub-types also have a weak or no secondary $i$-band maximum.  Using the time of primary $i$-band maximum  in conjunction with  the color stretch parameters, \sBV\ or \sgr,  it was found that normal,  2003fg-like, 1991bg-like,  and 2002cx-like SNe fall into distinct photometric groups which correspond to their spectral classification. 

We speculate that the timing of the $i$-band maximum is potentially caused by multiple factors: i) 1991bg-like SNe have 
low \Nifs\ masses, low ionization and an early onset of the recombination front, which merges the two $i$-band   maxima normally seen in SNe~Ia, ii)   2003fg-like SNe have a  lack of secondary $i$-band maximum which may be due to long diffusion time scales, caused by a large optically thick envelope around the \Nifs\ region, iii) and 2002cx-like SNe have a high ionization state, lack of recombination of iron group elements,  and low expansion velocities,  all of which produce a slow temperature evolution and an ejecta which slowly adiabatically cools. This may cause a lack of secondary $i$-band maximum, and a later primary maximum. 

Future surveys, such as LSST, will produce massive databases of high cadence light curves of SNe, and using the  identification scheme presented here it will be possible to unambiguously identify the SNe~Ia sub-type where there is poor, uncertain, or no spectral classification.  The results here can also be used in conjunction with already existing spectral  classifications, such as those from \citet{Branch06} to further refine our understanding of SNe~Ia physics.  This work can also be used as a guide line to screen out non-Ia SNe  from cosmological samples that don’t have complete spectroscopic typing. Finally, as more sub-types of thermonuclear SNe are discovered, high cadence $i$-band data can be used to determine their connection with the existing sub-types.




\vspace{0.5cm}
\acknowledgements

M.S. is supported in part by a generous grant (13261) from VILLUM FONDEN.
 E.B. acknowledges support from NASA Grant: NNX16AB25G.
 N.B.S. acknowledges support from  the Texas A\&M University Mitchell/Heep/Munnerlyn Chair in Observational Astronomy.
L.G. was funded by the European Union's Horizon 2020 research and innovation programme under the Marie Sk\l{}odowska-Curie grant agreement No. 839090.
The CSP-II has been funded by the NSF under grants AST-0306969, AST-0607438, AST-1008343, AST-1613426, AST-1613455, and AST-1613472, and in part by  a Sapere Aude Level 2 grant funded by the Danish Agency for Science and Technology and Innovation  (PI M.S.).

Facilities:  du Pont,  Swope, Python, IRAF, IDL

\acknowledgments

\appendix
\section{table}
Table \ref{table:table_params} contains the parameters derived the from light curves of the SNe. 

\startlongtable
\begin{deluxetable*}{lccccccc}
\tabletypesize{\small}
\centering
\tablecaption{Table of SNe parameters\label{table:table_params}}
\tablehead{\colhead{SN name} & \colhead{redshift} & \colhead{$t_{max}^{B}$} &  \colhead{$t_{max}^{i-B}$} & \colhead{$t_{max}^{i-g}$}  & \colhead{$S_{BV}$}  & \colhead{$S_{gr}$} & \colhead{SN type}\\
\colhead{}&\colhead{}&\colhead{MJD}&\colhead{days}&\colhead{days}}
\startdata
SN2006bt	&	0.032	&	53857.77	$\pm$	0.21	&	0.87	$\pm$	0.31	&	0.04	$\pm$	0.30	&	1.04	$\pm$	0.06	&	1.01	$\pm$	0.03	&	06bt-like	\\
SN2006ot	&	0.053	&	54065.68 $\downarrow$			&	2.8 $\uparrow$			&		$\cdots$		&	0.83 $\uparrow$			&		$\cdots$		&	06bt-like	\\ \hline
ASAS14kd	&	0.024	&	56981.76	$\pm$	0.21	&	$-$0.98	$\pm$	0.33	&		$\cdots$		&	1.13	$\pm$	0.04	&		$\cdots$		&	91T-like	\\
SN2014dl	&	0.033	&	56934.65	$\pm$	0.09	&	$-$1.31	$\pm$	0.14	&		$\cdots$		&	1.22	$\pm$	0.04	&		$\cdots$		&	91T-like	\\
SN2014eg	&	0.019	&	56991.26	$\pm$	0.47	&	$-$0.56	$\pm$	0.58	&		$\cdots$		&	1.17	$\pm$	0.04	&		$\cdots$		&	91T-like	\\
LSQ12gdj	&	0.030	&	56252.17	$\pm$	0.25	&	$-$1.65	$\pm$	0.27	&		$\cdots$		&	1.14	$\pm$	0.04	&		$\cdots$		&	91T-like	\\ \hline
SN2007if	&	0.074	&	54348.18	$\pm$	2.55	&	3.03	$\pm$	3.24	&	2.85	$\pm$	3.14	&	1.26	$\pm$	0.15	&	1.15	$\pm$	0.05	&	03fg-like	\\
LSQ14fmg	&	0.066	&	56938.71	$\pm$	0.58	&	1.51	$\pm$	1.05	&		$\cdots$		&	1.20	$\pm$	0.08	&		$\cdots$		&	03fg-like	\\
ASASSN-15hy	&	0.019	&	57150.67	$\pm$	0.33	&	7.28	$\pm$	0.47	&	5.97	$\pm$	0.65	&	1.24	$\pm$	0.18	&		$\cdots$		&	03fg-like	\\
SN2013ao	&	0.044	&	56362.11	$\pm$	0.11	&	1.80	$\pm$	0.26	&	3.24	$\pm$	0.69	&	1.02	$\pm$	0.13	&	1.09	$\pm$	0.04	&	03fg-like	\\
MLS140102\tablenotemark{a}	&	0.077	&	56667.36	$\pm$	0.31	&	3.97	$\pm$	0.73	&	   $\cdots$         &	1.24\tablenotemark{b} $\uparrow$		        &	$\cdots$	&	03fg-like	\\
ASASSN-15pz	&	0.000	&	57306.65	$\pm$	0.76	&	0.04	$\pm$	0.81	&		$\cdots$		&	1.37	$\pm$	0.09	&		$\cdots$		&	03fg-like	\\
SN2012dn	&	0.010	&	56132.59	$\pm$	0.59	&	2.04	$\pm$	0.61	&	$-$0.55	$\pm$	1.50	&	1.25	$\pm$	0.15	&	1.14	$\pm$	0.09	&	03fg-like	\\
SN2006gz	&	0.024	&	54019.97	$\pm$	0.30	&	4.60	$\pm$	1.65	&		$\cdots$		&	1.37	$\pm$	0.03	&		$\cdots$		&	03fg-like	\\
SN2009dc	&	0.021	&	54946.94	$\pm$	0.21	&	2.52	$\pm$	0.34	&	1.91	$\pm$	0.49	&	1.29	$\pm$	0.07	&		$\cdots$		&	03fg-like	\\ \hline
SN2012Z	&	0.007	&	55967.22	$\pm$	0.23	&	10.50	$\pm$	0.60	&	8.60	$\pm$	0.60	&	0.57	$\pm$	0.05	&	0.78	$\pm$	0.03	&	02cx-like	\\
PTF14ans	&	0.032	&	56781.76	$\pm$	0.31	&	10.85	$\pm$	0.65	&	10.18	$\pm$	0.58	&	0.70	$\pm$	0.08	&	0.83	$\pm$	0.03	&	02cx-like	\\
SN2014ek	&	0.023	&	56957.39	$\pm$	0.25	&	8.08	$\pm$	0.37	&	6.50	$\pm$	0.32	&	0.66	$\pm$	0.06	&	0.69	$\pm$	0.04	&	02cx-like	\\
SN2005hk	&	0.013	&	53684.62	$\pm$	0.12	&	9.94	$\pm$	0.25	&	9.16	$\pm$	0.26	&	0.77	$\pm$	0.06	&	0.77	$\pm$	0.04	&	02cx-like	\\
SN2008ha	&	0.005	&	54781.89	$\pm$	0.37	&	6.47	$\pm$	0.39	&	5.36	$\pm$	0.19	&	0.56	$\pm$	0.06	&	0.55	$\pm$	0.02	&	02cx-like	\\ \hline
LSQ14ip	&	0.061	&	56688.06	$\pm$	0.150	&	0.71	$\pm$	0.214	&		$\cdots$		&	0.50	$\pm$	0.041	&		$\cdots$		&	91bg-like	\\
KISS15m	&	0.024	&	57143.68	$\pm$	0.180	&	4.02	$\pm$	0.211	&	3.49	$\pm$	0.121	&	0.38	$\pm$	0.042	&	0.40	$\pm$	0.010	&	91bg-like	\\
SN2016hnk	&	0.016	&	57689.48	$\pm$	3.27	&	1.60	$\pm$	0.20	&		$\cdots$		&	0.44	$\pm$	0.03	&		$\cdots$		&	91bg-like	\\ 
SN2015bo	&	0.016	&	57075.76	$\pm$	0.110	&	2.05	$\pm$	0.140	&	1.64	$\pm$	0.093	&	0.47	$\pm$	0.040	&	0.54	$\pm$	0.008	&	91bg-like	\\
SN2009F	&	0.013	&	54841.80	$\pm$	0.110	&	3.48	$\pm$	0.176	&	3.01	$\pm$	0.140	&	0.33	$\pm$	0.041	&	0.43	$\pm$	0.011	&	91bg-like	\\
SN2008bt	&	0.015	&	54571.98	$\pm$	0.200	&	0.65	$\pm$	0.397	&	0.23	$\pm$	0.347	&	0.47	$\pm$	0.041	&	0.53	$\pm$	0.011	&	91bg-like	\\
SN2007N	&	0.013	&	54123.46	$\pm$	0.250	&	4.60	$\pm$	0.322	&	4.18	$\pm$	0.213	&	0.31	$\pm$	0.041	&	0.41	$\pm$	0.010	&	91bg-like	\\
SN2007ba	&	0.039	&	54196.67	$\pm$	0.130	&	0.48	$\pm$	0.230	&	$-$0.37	$\pm$	0.195	&	0.54	$\pm$	0.041	&	0.64	$\pm$	0.028	&	91bg-like	\\
LSQ11pn	&	0.033	&	55929.07	$\pm$	0.090	&	0.89	$\pm$	0.178	&	0.53	$\pm$	0.158	&	0.50	$\pm$	0.041	&	0.57	$\pm$	0.015	&	91bg-like	\\
SN2007ax	&	0.007	&	54187.75	$\pm$	0.110	&	4.17	$\pm$	0.210	&	3.63	$\pm$	0.180	&	0.36	$\pm$	0.041	&	0.42	$\pm$	0.011	&	91bg-like	\\
SN2006mr	&	0.006	&	54050.54	$\pm$	0.060	&	3.81	$\pm$	0.217	&	3.08	$\pm$	0.210	&	0.24	$\pm$	0.040	&	0.32	$\pm$	0.007	&	91bg-like	\\
SN2005ke	&	0.005	&	53698.39	$\pm$	0.080	&	2.80	$\pm$	0.205	&	1.95	$\pm$	0.190	&	0.42	$\pm$	0.040	&	0.48	$\pm$	0.007	&	91bg-like	\\ \hline
LSQ15alq	&	0.047	&	57154.13	$\pm$	0.20	&	$-$2.15	$\pm$	0.22	&		$\cdots$		&	0.91	$\pm$	0.04	&		$\cdots$		&	normal	\\
PS1-14rx	&	0.067	&	56736.32	$\pm$	0.29	&	$-$2.48	$\pm$	0.43	&		$\cdots$		&	0.83	$\pm$	0.04	&		$\cdots$		&	normal	\\
PTF11pbp	&	0.029	&	55871.42	$\pm$	0.18	&	$-$2.09	$\pm$	0.23	&		$\cdots$		&	1.19	$\pm$	0.04	&		$\cdots$		&	normal	\\
PTF13anh	&	0.062	&	56413.56	$\pm$	0.16	&	$-$1.56	$\pm$	0.23	&		$\cdots$		&	0.96	$\pm$	0.04	&		$\cdots$		&	normal	\\
PTF13duj	&	0.017	&	56601.23	$\pm$	0.19	&	$-$0.93	$\pm$	0.31	&		$\cdots$		&	1.19	$\pm$	0.04	&		$\cdots$		&	normal	\\
PTF13ebh	&	0.013	&	56622.90	$\pm$	0.08	&	$-$1.30	$\pm$	0.11	&	$-$1.83	$\pm$	0.09	&	0.61	$\pm$	0.04	&	0.65	$\pm$	0.01	&	normal	\\
PTF13efe	&	0.075	&	56640.64	$\pm$	0.31	&	$-$1.22	$\pm$	0.46	&		$\cdots$		&	1.23	$\pm$	0.04	&		$\cdots$		&	normal	\\
PTF14aaf	&	0.059	&	56739.46	$\pm$	0.43	&	$-$0.13	$\pm$	0.49	&		$\cdots$		&	1.19	$\pm$	0.04	&		$\cdots$		&	normal	\\
PTF14aje	&	0.028	&	56757.99	$\pm$	0.27	&	$-$1.26	$\pm$	0.31	&	$-$1.88	$\pm$	0.18	&	0.68	$\pm$	0.04	&	0.71	$\pm$	0.02	&	normal	\\
PTF14fpg	&	0.034	&	56930.56	$\pm$	0.27	&	$-$2.20	$\pm$	0.35	&	$-$2.89	$\pm$	0.24	&	1.21	$\pm$	0.04	&	1.17	$\pm$	0.04	&	normal	\\
LSQ14xi	&	0.051	&	56719.97	$\pm$	0.31	&	$-$2.04	$\pm$	0.37	&		$\cdots$		&	1.10	$\pm$	0.04	&		$\cdots$		&	normal	\\
LSQ14wp	&	0.070	&	56716.98	$\pm$	0.19	&	$-$2.81	$\pm$	0.49	&	$-$2.72	$\pm$	0.46	&	1.12	$\pm$	0.04	&	1.17	$\pm$	0.02	&	normal	\\
LSQ14mc	&	0.057	&	56696.88	$\pm$	0.21	&	$-$3.62	$\pm$	0.26	&		$\cdots$		&	0.95	$\pm$	0.04	&		$\cdots$		&	normal	\\
PTF14gnl	&	0.054	&	56956.29	$\pm$	0.41	&	$-$2.91	$\pm$	0.40	&		$\cdots$		&	1.04	$\pm$	0.04	&		$\cdots$		&	normal	\\
LSQ13dsm	&	0.042	&	56670.15	$\pm$	0.26	&	$-$2.69	$\pm$	0.27	&	$-$2.59	$\pm$	0.11	&	0.88	$\pm$	0.04	&	0.94	$\pm$	0.02	&	normal	\\
LSQ13ry	&	0.030	&	56395.38	$\pm$	0.18	&	$-$2.84	$\pm$	0.20	&	$-$2.91	$\pm$	0.09	&	0.86	$\pm$	0.04	&	0.93	$\pm$	0.02	&	normal	\\
LSQ14age	&	0.081	&	56734.76	$\pm$	0.49	&	$-$0.68	$\pm$	0.53	&		$\cdots$		&	1.12	$\pm$	0.05	&		$\cdots$		&	normal	\\
LSQ14q	&	0.067	&	56672.60	$\pm$	0.29	&	$-$2.21	$\pm$	0.31	&		$\cdots$		&	0.94	$\pm$	0.04	&		$\cdots$		&	normal	\\
LSQ14ahm	&	0.050	&	56736.19	$\pm$	0.45	&	$-$2.29	$\pm$	0.47	&		$\cdots$		&	1.23	$\pm$	0.04	&		$\cdots$		&	normal	\\
LSQ14foj	&	0.046	&	56939.65	$\pm$	0.25	&	$-$0.99	$\pm$	0.30	&		$\cdots$		&	1.05	$\pm$	0.04	&		$\cdots$		&	normal	\\
LSQ14fom	&	0.056	&	56937.59	$\pm$	0.16	&	$-$2.25	$\pm$	0.24	&		$\cdots$		&	0.92	$\pm$	0.04	&		$\cdots$		&	normal	\\
LSQ14gov	&	0.090	&	57027.74	$\pm$	0.18	&	$-$2.58	$\pm$	0.29	&		$\cdots$		&	1.11	$\pm$	0.04	&		$\cdots$		&	normal	\\
LSQ14ie	&	0.090	&	56687.97	$\pm$	0.30	&	$-$2.09	$\pm$	0.50	&		$\cdots$		&	1.22	$\pm$	0.04	&		$\cdots$		&	normal	\\
SN2005W	&	0.009	&	53411.33	$\pm$	0.25	&	$-$2.29	$\pm$	0.27	&		$\cdots$		&	0.92	$\pm$	0.04	&		$\cdots$		&	normal	\\
LSQ14jp	&	0.045	&	56692.21	$\pm$	0.11	&	$-$0.49	$\pm$	0.13	&		$\cdots$		&	0.64	$\pm$	0.04	&		$\cdots$		&	normal	\\
SN2007bd	&	0.031	&	54206.29	$\pm$	0.24	&	$-$3.53	$\pm$	0.31	&	$-$3.98	$\pm$	0.21	&	0.88	$\pm$	0.04	&	0.97	$\pm$	0.03	&	normal	\\
LSq13dpm	&	0.051	&	56656.91	$\pm$	0.15	&	$-$2.76	$\pm$	0.25	&		$\cdots$		&	1.07	$\pm$	0.04	&		$\cdots$		&	normal	\\
SN2004eo	&	0.016	&	53278.24	$\pm$	0.11	&	$-$2.90	$\pm$	0.12	&	$-$3.50	$\pm$	0.08	&	0.82	$\pm$	0.04	&	0.87	$\pm$	0.01	&	normal	\\
SN2004ey	&	0.016	&	53303.73	$\pm$	0.13	&	$-$2.98	$\pm$	0.16	&	$-$3.37	$\pm$	0.10	&	1.01	$\pm$	0.04	&	1.09	$\pm$	0.01	&	normal	\\
SN2004gs	&	0.027	&	53355.92	$\pm$	0.13	&	$-$1.90	$\pm$	0.40	&	$-$2.28	$\pm$	0.38	&	0.70	$\pm$	0.04	&	0.74	$\pm$	0.01	&	normal	\\
SN2005el	&	0.015	&	53647.02	$\pm$	0.45	&	$-$3.04	$\pm$	0.49	&	$-$2.86	$\pm$	0.20	&	0.84	$\pm$	0.04	&	0.87	$\pm$	0.01	&	normal	\\
SN2005iq	&	0.034	&	53687.36	$\pm$	0.05	&	$-$3.14	$\pm$	0.12	&	$-$3.37	$\pm$	0.11	&	0.87	$\pm$	0.04	&	0.96	$\pm$	0.04	&	normal	\\
SN2005kc	&	0.015	&	53697.61	$\pm$	0.16	&	$-$3.05	$\pm$	0.21	&		$\cdots$		&	0.90	$\pm$	0.04	&		$\cdots$		&	normal	\\
SN2015F	&	0.005	&	57106.46	$\pm$	0.26	&	$-$2.47	$\pm$	0.31	&	$-$2.88	$\pm$	0.18	&	0.85	$\pm$	0.04	&	0.9	$\pm$	0.01	&	normal	\\
SN2014I	&	0.030	&	56683.53	$\pm$	0.14	&	$-$2.65	$\pm$	0.17	&	$-$2.69	$\pm$	0.11	&	0.90	$\pm$	0.04	&	0.96	$\pm$	0.02	&	normal	\\
SN2005ki	&	0.019	&	53704.75	$\pm$	0.14	&	$-$2.97	$\pm$	0.27	&	$-$3.67	$\pm$	0.24	&	0.83	$\pm$	0.04	&	0.86	$\pm$	0.01	&	normal	\\
SN2014at	&	0.032	&	56774.07	$\pm$	0.17	&	$-$3.44	$\pm$	0.32	&	$-$3.29	$\pm$	0.27	&	0.93	$\pm$	0.04	&	1.02	$\pm$	0.02	&	normal	\\
SN2013M	&	0.035	&	56323.28	$\pm$	0.28	&	$-$1.89	$\pm$	0.33	&	$-$2.00	$\pm$	0.20	&	0.93	$\pm$	0.04	&	0.96	$\pm$	0.02	&	normal	\\
SN2013H	&	0.016	&	56309.43	$\pm$	0.18	&	$-$2.99	$\pm$	0.20	&	$-$3.34	$\pm$	0.11	&	1.06	$\pm$	0.04	&	1.12	$\pm$	0.01	&	normal	\\
PTF14w	&	0.019	&	56669.55	$\pm$	0.17	&	$-$2.21	$\pm$	0.19	&	$-$2.59	$\pm$	0.10	&	0.73	$\pm$	0.04	&	0.75	$\pm$	0.01	&	normal	\\
SN2013gy	&	0.014	&	56648.71	$\pm$	0.15	&	$-$2.89	$\pm$	0.16	&	$-$2.96	$\pm$	0.09	&	0.89	$\pm$	0.04	&	0.96	$\pm$	0.01	&	normal	\\
SN2013E	&	0.009	&	56307.38	$\pm$	0.22	&	$-$2.54	$\pm$	0.24	&	$-$3.40	$\pm$	0.10	&	1.12	$\pm$	0.04	&	1.09	$\pm$	0.01	&	normal	\\
SN2005M	&	0.022	&	53405.50	$\pm$	0.19	&	$-$2.93	$\pm$	0.21	&	$-$3.03	$\pm$	0.09	&	1.21	$\pm$	0.04	&	1.18	$\pm$	0.02	&	normal	\\
SN2013aj	&	0.009	&	56361.45	$\pm$	0.27	&	$-$3.42	$\pm$	0.39	&	$-$3.46	$\pm$	0.28	&	0.78	$\pm$	0.04	&	0.83	$\pm$	0.01	&	normal	\\
SN2013aa	&	0.004	&	56343.15	$\pm$	0.20	&	$-$3.63	$\pm$	0.34	&	$-$3.76	$\pm$	0.28	&	1.00	$\pm$	0.04	&	1.1	$\pm$	0.01	&	normal	\\
SN2012ij	&	0.011	&	56302.33	$\pm$	0.10	&	$-$0.06	$\pm$	0.16	&	$-$0.54	$\pm$	0.13	&	0.53	$\pm$	0.04	&	0.6	$\pm$	0.01	&	normal	\\
SN2012ht	&	0.004	&	56295.22	$\pm$	0.39	&	$-$1.81	$\pm$	0.42	&	$-$2.18	$\pm$	0.19	&	0.86	$\pm$	0.04	&	0.92	$\pm$	0.02	&	normal	\\
SN2012hr	&	0.008	&	56288.72	$\pm$	0.22	&	$-$3.04	$\pm$	0.23	&	$-$3.28	$\pm$	0.09	&	0.96	$\pm$	0.04	&	1.02	$\pm$	0.01	&	normal	\\
SN2012hd	&	0.012	&	56264.97	$\pm$	0.14	&	$-$2.44	$\pm$	0.18	&	$-$2.82	$\pm$	0.11	&	0.87	$\pm$	0.04	&	0.92	$\pm$	0.01	&	normal	\\
SN2012gm	&	0.015	&	56261.79	$\pm$	0.08	&	$-$3.29	$\pm$	0.11	&		$\cdots$		&	0.98	$\pm$	0.04	&		$\cdots$		&	normal	\\
SN2012fr	&	0.006	&	56242.28	$\pm$	0.34	&	$-$2.67	$\pm$	0.42	&	$-$2.91	$\pm$	0.26	&	1.12	$\pm$	0.04	&	1.19	$\pm$	0.01	&	normal	\\
SN2011jh	&	0.008	&	55930.21	$\pm$	0.12	&	$-$2.16	$\pm$	0.13	&	$-$2.93	$\pm$	0.07	&	0.80	$\pm$	0.04	&	0.84	$\pm$	0.01	&	normal	\\
PTF14yy	&	0.043	&	56732.58	$\pm$	0.19	&	$-$0.87	$\pm$	0.24	&		$\cdots$		&	0.82	$\pm$	0.04	&		$\cdots$		&	normal	\\
SN2013fy	&	0.031	&	56600.55	$\pm$	0.22	&	$-$0.24	$\pm$	0.28	&		$\cdots$		&	1.19	$\pm$	0.04	&		$\cdots$		&	normal	\\
SN2006gj	&	0.028	&	54000.06	$\pm$	0.22	&	$-$0.91	$\pm$	0.43	&	$-$1.44	$\pm$	0.38	&	0.66	$\pm$	0.04	&	0.88	$\pm$	0.07	&	normal	\\
ASAS14jz	&	0.016	&	56980.56	$\pm$	0.08	&	$-$2.79	$\pm$	0.14	&		$\cdots$		&	0.98	$\pm$	0.04	&		$\cdots$		&	normal	\\
ASAS14hu	&	0.022	&	56935.05	$\pm$	0.20	&	$-$2.60	$\pm$	0.29	&	$-$2.58	$\pm$	0.22	&	1.08	$\pm$	0.04	&	1.16	$\pm$	0.03	&	normal	\\
ASAS14hr	&	0.034	&	56932.12	$\pm$	0.21	&	$-$1.03	$\pm$	0.29	&	$-$1.21	$\pm$	0.22	&	0.78	$\pm$	0.04	&	0.85	$\pm$	0.02	&	normal	\\
ASAS14ad	&	0.026	&	56691.61	$\pm$	0.15	&	$-$2.12	$\pm$	0.18	&	$-$2.89	$\pm$	0.11	&	1.03	$\pm$	0.04	&	1.08	$\pm$	0.02	&	normal	\\
SN2006gt	&	0.045	&	54003.00	$\pm$	0.20	&	$-$0.81	$\pm$	0.24	&	$-$1.23	$\pm$	0.15	&	0.56	$\pm$	0.04	&	0.68	$\pm$	0.03	&	normal	\\
SN2006kf	&	0.021	&	54041.03	$\pm$	0.16	&	$-$2.43	$\pm$	0.19	&	$-$2.81	$\pm$	0.12	&	0.74	$\pm$	0.04	&	0.77	$\pm$	0.01	&	normal	\\
SN2009Y	&	0.009	&	54875.37	$\pm$	0.20	&	$-$1.51	$\pm$	0.27	&	$-$2.57	$\pm$	0.20	&	1.19	$\pm$	0.04	&	1.02	$\pm$	0.01	&	normal	\\
SN2009ds	&	0.019	&	54960.03	$\pm$	0.16	&	$-$1.90	$\pm$	0.33	&		$\cdots$		&	1.13	$\pm$	0.05	&		$\cdots$		&	normal	\\
SN2009D	&	0.025	&	54840.61	$\pm$	0.32	&	$-$3.04	$\pm$	0.35	&	$-$3.20	$\pm$	0.15	&	1.19	$\pm$	0.04	&	1.16	$\pm$	0.03	&	normal	\\
SN2009cz	&	0.021	&	54942.31	$\pm$	0.18	&	$-$2.82	$\pm$	0.22	&		$\cdots$		&	1.19	$\pm$	0.04	&		$\cdots$		&	normal	\\
SN2009ad	&	0.028	&	54885.59	$\pm$	0.16	&	$-$2.87	$\pm$	0.18	&		$\cdots$		&	1.02	$\pm$	0.04	&		$\cdots$		&	normal	\\
SN2009ab	&	0.011	&	54882.94	$\pm$	0.19	&	$-$3.15	$\pm$	0.20	&		$\cdots$		&	0.87	$\pm$	0.04	&		$\cdots$		&	normal	\\
SN2007ca	&	0.014	&	54226.95	$\pm$	0.25	&	$-$2.94	$\pm$	0.33	&		$\cdots$		&	1.06	$\pm$	0.04	&		$\cdots$		&	normal	\\
SN2007le	&	0.007	&	54398.40	$\pm$	0.36	&	$-$2.19	$\pm$	0.39	&	$-$2.80	$\pm$	0.16	&	1.03	$\pm$	0.04	&	1.05	$\pm$	0.01	&	normal	\\
SN2007af	&	0.006	&	54174.07	$\pm$	0.24	&	$-$2.84	$\pm$	0.28	&	$-$3.12	$\pm$	0.14	&	0.93	$\pm$	0.04	&	1.0	$\pm$	0.01	&	normal	\\
SN2007on	&	0.007	&	54420.16	$\pm$	0.09	&	$-$1.99	$\pm$	0.22	&	$-$2.03	$\pm$	0.20	&	0.57	$\pm$	0.04	&	0.62	$\pm$	0.01	&	normal	\\
SN2008bf	&	0.024	&	54554.46	$\pm$	0.22	&	$-$3.74	$\pm$	0.28	&	$-$3.81	$\pm$	0.18	&	1.02	$\pm$	0.04	&	1.12	$\pm$	0.02	&	normal	\\
SN2006ob	&	0.059	&	54062.95	$\pm$	0.20	&	$-$1.20	$\pm$	0.41	&	$-$1.69	$\pm$	0.36	&	0.72	$\pm$	0.04	&	0.79	$\pm$	0.03	&	normal	\\
SN2008fp	&	0.006	&	54729.53	$\pm$	0.29	&	$-$2.25	$\pm$	0.31	&	$-$2.91	$\pm$	0.11	&	1.08	$\pm$	0.04	&	1.12	$\pm$	0.01	&	normal	\\
SN2008gp	&	0.033	&	54778.82	$\pm$	0.24	&	$-$3.67	$\pm$	0.26	&	$-$3.74	$\pm$	0.13	&	0.97	$\pm$	0.04	&	1.03	$\pm$	0.08	&	normal	\\
SN2008hj	&	0.038	&	54800.83	$\pm$	0.14	&	$-$2.79	$\pm$	0.19	&		$\cdots$		&	1.01	$\pm$	0.04	&		$\cdots$		&	normal	\\
SN2008hv	&	0.013	&	54816.74	$\pm$	0.28	&	$-$3.53	$\pm$	0.30	&	$-$4.00	$\pm$	0.11	&	0.85	$\pm$	0.04	&	0.89	$\pm$	0.02	&	normal	\\
SN2008bc	&	0.015	&	54548.83	$\pm$	0.17	&	$-$2.66	$\pm$	0.21	&	$-$3.28	$\pm$	0.13	&	1.05	$\pm$	0.04	&	1.08	$\pm$	0.01	&	normal	\\
LSQ13dkp	&	0.069	&	56641.97	$\pm$	0.14	&	$-$1.71	$\pm$	0.29	&		$\cdots$		&	0.63	$\pm$	0.04	&		$\cdots$		&	normal	\\
ASAS14me	&	0.018	&	57019.67	$\pm$	0.19	&	$-$2.92	$\pm$	0.20	&	$-$2.88	$\pm$	0.09	&	1.09	$\pm$	0.04	&	1.19	$\pm$	0.04	&	normal	\\
LSQ13cwp	&	0.067	&	56611.57	$\pm$	0.32	&	$-$0.75	$\pm$	0.41	&		$\cdots$		&	0.91	$\pm$	0.05	&		$\cdots$		&	normal	\\
LSQ13aiz	&	0.009	&	56437.25	$\pm$	0.14	&	$-$0.82	$\pm$	0.21	&		$\cdots$		&	0.96	$\pm$	0.04	&		$\cdots$		&	normal	\\
LSQ12hzs	&	0.072	&	56298.56	$\pm$	0.16	&	$-$2.12	$\pm$	0.27	&		$\cdots$		&	1.09	$\pm$	0.05	&		$\cdots$		&	normal	\\
LSQ12gxj	&	0.036	&	56274.88	$\pm$	0.17	&	$-$2.36	$\pm$	0.23	&		$\cdots$		&	1.10	$\pm$	0.04	&		$\cdots$		&	normal	\\
SN2006ax	&	0.017	&	53827.11	$\pm$	0.25	&	$-$3.26	$\pm$	0.29	&	$-$3.06	$\pm$	0.16	&	0.99	$\pm$	0.04	&	1.07	$\pm$	0.02	&	normal	\\
LSQ12fxd	&	0.031	&	56246.07	$\pm$	0.19	&	$-$2.52	$\pm$	0.31	&		$\cdots$		&	1.10	$\pm$	0.04	&		$\cdots$		&	normal	\\
LSQ12fvl	&	0.056	&	56239.83	$\pm$	0.11	&	0.00	$\pm$	0.22	&		$\cdots$		&	0.59	$\pm$	0.04	&		$\cdots$		&	normal	\\
LSQ12bld	&	0.083	&	56025.04	$\pm$	0.32	&	$-$2.09	$\pm$	0.36	&		$\cdots$		&	0.85	$\pm$	0.04	&		$\cdots$		&	normal	\\
SN2006bh	&	0.011	&	53833.31	$\pm$	0.11	&	$-$3.59	$\pm$	0.23	&	$-$3.72	$\pm$	0.20	&	0.80	$\pm$	0.04	&	0.86	$\pm$	0.01	&	normal	\\
SN2006D	&	0.009	&	53757.35	$\pm$	0.14	&	$-$2.56	$\pm$	0.19	&	$-$2.83	$\pm$	0.13	&	0.82	$\pm$	0.04	&	0.84	$\pm$	0.01	&	normal	\\
KISS13v	&	0.080	&	56395.62	$\pm$	0.13	&	$-$0.23	$\pm$	0.31	&		$\cdots$		&	1.03	$\pm$	0.05	&		$\cdots$		&	normal	\\
PSN 13-24\tablenotemark{c}	&	0.020	&	57069.77	$\pm$	0.21	&	$-$0.77	$\pm$	0.51	&	$-$1.14	$\pm$	0.47	&	0.71	$\pm$	0.04	&	0.76	$\pm$	0.02	&	normal	\\
SN2006et	&	0.022	&	53993.16	$\pm$	0.14	&	$-$2.68	$\pm$	0.18	&	$-$2.98	$\pm$	0.12	&	1.09	$\pm$	0.04	&	1.15	$\pm$	0.02	&	normal	\\
ASAS14mw	&	0.027	&	57028.20	$\pm$	0.18	&	$-$3.05	$\pm$	0.42	&	$-$3.83	$\pm$	0.38	&	1.07	$\pm$	0.04	&	1.11	$\pm$	0.01	&	normal	\\
ASAS14my	&	0.021	&	57029.99	$\pm$	0.27	&	$-$2.08	$\pm$	0.30	&	$-$2.82	$\pm$	0.16	&	0.91	$\pm$	0.04	&	0.98	$\pm$	0.02	&	normal	\\
ASAS15be	&	0.022	&	57048.59	$\pm$	0.29	&	$-$1.42	$\pm$	0.44	&	$-$2.23	$\pm$	0.34	&	1.20	$\pm$	0.04	&	1.14	$\pm$	0.01	&	normal	\\
ASAS15bm	&	0.021	&	57053.21	$\pm$	0.31	&	$-$2.26	$\pm$	0.37	&	$-$3.11	$\pm$	0.22	&	0.99	$\pm$	0.04	&	1.11	$\pm$	0.08	&	normal	\\
ASAS14lw	&	0.021	&	57012.25	$\pm$	0.40	&	$-$2.26	$\pm$	0.41	&		$\cdots$		&	1.26	$\pm$	0.04	&		$\cdots$		&	normal	\\
CSS120305\tablenotemark{d}	&	0.097	&	56022.06	$\pm$	0.20	&	$-$2.42	$\pm$	0.34	&		$\cdots$		&	1.11	$\pm$	0.04	&		$\cdots$		&	normal	\\
CSS130303\tablenotemark{e}	&	0.079	&	56364.02	$\pm$	0.38	&	$-$1.57	$\pm$	0.43	&		$\cdots$		&	1.19	$\pm$	0.04	&		$\cdots$		&	normal	\\
OGLE14-019\tablenotemark{f}	&	0.036	&	56718.28	$\pm$	0.25	&	$-$2.06	$\pm$	0.34	&	$-$1.67	$\pm$	0.28	&	0.89	$\pm$	0.04	&	1.12	$\pm$	0.03	&	normal	\\
ASAS15hf	&	0.006	&	57136.89	$\pm$	0.13	&	$-$2.67	$\pm$	0.23	&	$-$3.26	$\pm$	0.19	&	0.92	$\pm$	0.04	&	0.89	$\pm$	0.01	&	normal	\\
SN2004ef	&	0.031	&	53263.77	$\pm$	0.12	&	$-$3.32	$\pm$	0.12	&	$-$3.70	$\pm$	0.04	&	0.82	$\pm$	0.04	&	0.83	$\pm$	0.01	&	normal	\\
\enddata
\tablenotetext{a}{SN full name: MLS140102:120307-010132}
\tablenotetext{b}{This lower limit was calculated using the \sBV\ vs $\Delta \rm{m}_{15}($$B$$)$ relation for the 03fg-like SNe.}
\tablenotetext{c}{SN full name: PSN J13471211-2422171}
\tablenotetext{d}{SN full name: CSS120325:123816-150632}
\tablenotetext{e}{SN full name: CSS130303:105206-133424}
\tablenotetext{f}{SN full name: OGLE-2014-SN-019}
\end{deluxetable*}

\end{document}